\documentclass[a4paper]{PoS}
\usepackage{amsmath}
\usepackage{amssymb}
\usepackage{epsfig}
\usepackage{euscript}
\def\mathswitchr#1{\relax\ifmmode{\mathrm{#1}}\else$\mathrm{#1}$\fi}

\newcommand {\pslash}{\hbox{$\not\hbox{\kern-2.3pt $p$}$}}

\usepackage{cite}

\usepackage{epic}

\def\alf1{ {\alpha\over\pi} }
\def\rQCED{{\rm QCED}}
 
\title{Comparisons of Predictions from Exact Amplitude-Based Resummation Methods with LHC and Cosmological Data}
\ShortTitle{Comparisons of Predictions from Exact Amplitude-Based Resummation Methods with LHC and Cosmological Data}
\author{\speaker{B.F.L. Ward}%
    \thanks{Work supported in part by D.o.E. grant DE-FG02-09ER41600.}\\
      Department of Physics, Baylor University\\
        E-mail:\email{bfl\_ward@baylor.edu}}
\author{S.K. Majhi\\
      Department of Theoretical Physics, Indian Association for the Cultivation of Science\\
        E-mail: \email{tpskm@iacs.res.in}}
\author{A. Mukhopadhyay \\
      Department of Physics, Baylor University\\
        E-mail: \email{Aditi\_Mukhopadhyay@baylor.edu}}
\author{S.A. Yost
 \thanks{Work supported in part by 
D.o.E. grant DE-PS02-09ER09-01 and grants from The Citadel Foundation.}\\
  Department of Physics, The Citadel\\
        E-mail: \email{scott.yost@citadel.edu}}
\abstract{
We present the current status of the comparisons with the respective data of the predictions of our approach of exact amplitude-based resummation in quantum field theory in two areas of investigation: precision QCD calculations of all four of us as needed for LHC physics and the resummed quantum gravity realization by one of us (B.F.L.W.) of Einstein's theory of general relativity as formulated by Feynman. The agreement between the theoretical predictions and the data exhibited continues to be encouraging.}

\FullConference{11th International Symposium on Radiative Corrections (Applications of Quantum Field Theory to Phenomenology)\\
                 22-27 September 2013\\
                 Lumley Castle Hotel, Durham, UK\\
          BU-HEPP-13-04,\\ 
             Dec., 2013 }

\begin{document}
%

 
\baselineskip=12pt 
\def\Kmax{K_{\rm max}}\def\ieps{{i\epsilon}}\def\rQCD{{\rm QCD}}
\section{\bf Introduction}\label{intro}\par
Successful running of the LHC during 2010-2012 has allowed the accumulation of large samples of data on SM standard candle processes such as heavy gauge boson production and decay to lepton pairs: for example, samples exceeding $10^7$
events for $Z/\gamma^*$ production and decay to $\ell\bar\ell,\; \ell=e,\mu,$ now exist for ATLAS and CMS. Such data emphasize that the era of precision QCD, wherein one needs
predictions for QCD processes at the total precision tag of $1\%$ or better,
has arrived. Its arrival makes more manifest the need for exact, 
amplitude-based resummation of large higher order effects; for, with 
such resummation one may have better than 1\% precision 
as a realistic goal as we shall show in what follows. Such precision allows
one to  
distinguish new physics(NP) from higher order SM processes and to distinguish 
different models of new physics from one another as well. 
In an analogous development, one of us(B.F.L.W.) has shown that the extension 
of the exact amplitude-based resummation approach to Einstein's 
theory of general relativity allows one to make contact with UV sensitive
cosmological data using
ordinary quantum field theoretic methods. 
Here, we present the status of these two 
applications of exact amplitude-based
resummation theory in quantum field theory in relation to recent available data
from the LHC and from cosmological observations.\par
Our discussion proceeds as follows.
First, we review the elements our approach to precision LHC physics, an amplitude-based QED$\otimes$QCD($\equiv\text{QCD}\otimes\text{QED}$) exact resummation 
theory~\cite{qced} 
realized by MC methods. We start from the
well-known 
fully differential representation
\begin{equation}
d\sigma =\sum_{i,j}\int dx_1dx_2F_i(x_1)F_j(x_2)d\hat\sigma_{\text{res}}(x_1x_2s)
\label{bscfrla}
\end{equation}
of a hard LHC scattering process, where $\{F_j\}$ and 
$d\hat\sigma_{\text{res}}$ are the respective parton densities and 
reduced hard differential cross section and we indicate that the latter 
has been resummed
for all large EW and QCD higher order corrections in a manner consistent
with achieving a total precision tag of 1\% or better for the total 
theoretical precision of (\ref{bscfrla}). 
The determination of the 
total theoretical precision $\Delta\sigma_{\text{th}}$ of (\ref{bscfrla}) is central to precision QCD theory. It can be decomposed into
its physical and technical components as defined in Refs.~\cite{jadach-prec,radcr11}. Knowledge of 
$\Delta\sigma_{\text{th}}$ is essential to the faithful application
of any theoretical prediction to precision experimental data for new physics signals, SM backgrounds, and over-all normalization considerations.
In general, if $\Delta\sigma_{\text{th}}\leq f\Delta\sigma_{\text{expt}}$, 
where $\Delta\sigma_{\text{expt}}$ is the respective experimental error
and $f\lesssim \frac{1}{2}$,
the theoretical uncertainty will not significantly affect the 
analysis of the data for physics studies in an adverse way. It was
with the goal of achieving such a provable theoretical precision tag that 
we have 
developed the $\text{QCD}\otimes\text{QED}$ resummation theory in Refs.~\cite{qced}
for all components of (\ref{bscfrla}).
The master formula for the starting point in all cases is  
{\small
\begin{eqnarray}
&d\bar\sigma_{\rm res} = e^{\rm SUM_{IR}(QCED)}
   \sum_{{n,m}=0}^\infty\frac{1}{n!m!}\int\prod_{j_1=1}^n\frac{d^3k_{j_1}}{k_{j_1}} \cr
&\prod_{j_2=1}^m\frac{d^3{k'}_{j_2}}{{k'}_{j_2}}
\int\frac{d^4y}{(2\pi)^4}e^{iy\cdot(p_1+q_1-p_2-q_2-\sum k_{j_1}-\sum {k'}_{j_2})+
D_\rQCED} \cr
&\tilde{\bar\beta}_{n,m}(k_1,\ldots,k_n;k'_1,\ldots,k'_m)\frac{d^3p_2}{p_2^{\,0}}\frac{d^3q_2}{q_2^{\,0}},
\label{subp15b}
\end{eqnarray}}\noindent
where $d\bar\sigma_{\rm res}$ is either the reduced cross section
$d\hat\sigma_{\rm res}$ or the differential rate associated to a
DGLAP-CS~\cite{dglap,cs} kernel involved in the evolution of the $\{F_j\}$ and 
where the {\em new} (YFS-style~\cite{yfs,sjbw}) {\em non-Abelian} residuals 
$\tilde{\bar\beta}_{n,m}(k_1,\ldots,k_n;k'_1,\ldots,k'_m)$ have $n$ hard gluons and $m$ hard photons and we show the generic $2f$ final state 
with momenta $p_2,\; q_2$ for
definiteness. The infrared functions ${\rm SUM_{IR}(QCED)}$, $ D_\rQCED\; $
are given in Refs.~\cite{qced,irdglap1,irdglap2}. The result (\ref{subp15b}) 
is exact and
its residuals $\tilde{\bar\beta}_{n,m}$ allow a rigorous parton 
shower/ME matching via their shower-subtracted 
counterparts $\hat{\tilde{\bar\beta}}_{n,m}$~\cite{qced}.\par

Using the result (\ref{subp15b}), one of us(B.F.L.W.) has shown in Ref.~\cite{rqg} that an exact, amplitude-based resummation approach to Feynman's formulation of Einstein's theory
is possible via
the following representation of the Feynman propagators in that theory: 
\begin{equation}
\begin{split}
i\Delta'_F(k)& =  \frac{i}{(k^2-m^2-\Sigma_s+i\epsilon)}\\
&=\frac{ie^{B''_g(k)}}{(k^2-m^2-\Sigma'_s+i\epsilon)}\\
&\equiv i\Delta'_F(k)|_{\text{resummed}}.
\end{split}
\label{rqg1}
\end{equation}
for scalar fields with an attendant generalization for spinning fields~\cite{rqg}. We stress that (\ref{rqg1}) is exact. $B''_g(k)$ is given in Refs.~\cite{rqg} and is presented below.\par
We now discuss in turn the two paradigms opened by (\ref{subp15b}) for precision QCD for the LHC and for exact resummation of Einstein's theory in the context of comparisons with recent data.\par

\section{Precision QCD for the LHC in Comparison to Data}
We first recall that, as we have shown in Refs.~\cite{herwiri1}, the methods we employ for resummation of the QCD theory
are fully consistent with the methods in Refs.~\cite{stercattrent1,scet1}. 
A key difference between our approach and the two in 
Refs.~\cite{stercattrent1,scet1} is that our approach is exact whereas
the latter approaches are approximate: in Refs.~\cite{stercattrent1}, the observation that, for any integrable function $f(z)$, 
$$|\int_0^1dz z^{n-1}f(z)|\le (\frac{1}{n})\max_{z\in[0,1]}\{|f(z)|\}$$ is used to drop 
non-singular terms in the cross section at $z\rightarrow 1$, the respective threshold point, in going to n-Mellin space to resum the respective large threshold effects; in Refs.~\cite{scet1}, terms of ${\cal O}(\lambda)$ for $\lambda=\sqrt{\Lambda/Q}$ are dropped, where $\lambda\sim 0.3\text{GeV}$ and $Q\sim 100\text{GeV}$, so that $\lambda\simeq 5.5\%$. The known equivalence of the 
two approaches in Refs.~\cite{stercattrent1,scet1} implies that a similar error holds for the approach in  Refs.~\cite{stercattrent1}. These two approaches may be used to construct approximations to our residuals $\tilde{\bar\beta}_{n,0}$. We will pursue such approximations elsewhere~\cite{elswh}.\par  
Similarly, we show in the fourth paper in Refs.~\cite{herwiri1} that the approach to resummation in Refs.~\cite{css}, which is realized in the  MC integration program R{\scriptsize ES}B{\scriptsize OS}~\cite{resbos} and which, for the case  of heavy gauge boson production in hadron colliders is presented from the fourth paper in Refs.~\cite{css} as
{\small
\begin{align}
\frac{d\sigma}{dQ^2dydQ_T^2}&\sim \frac{4\pi^2\alpha^2}{9Q^2s}\Bigg\{\int \frac{d^2b}{(2\pi)^2} e^{i\vec{Q}_T\cdot\vec{b}}\sum_je_j^2\widetilde{W}_j(b^*;Q,x_A,x_B)e^{\tiny\{-\ln(Q^2/Q_0^2)g_1(b)-g_{j/A}(x_A,b)-g_{j/B}(x_B,b)\}} \nonumber\\
&\qquad\qquad\qquad\qquad\qquad +\; Y(Q_T;Q,x_A,x_B)\Bigg\},
\label{css4} 
\end{align}}\noindent
is also approximate at the several \% level,  
where we have the usual kinematics so that $\vec{Q}_T=\vec{p}_T$ is the $\gamma*$ transverse momentum, A,B are protons at the LHC, $s$ is the cms squared energy of the protons, $ Q^{\mu}$ is the $\gamma*$ 4-momentum so that $Q^2$ is the $\gamma*$ mass squared, and $y=\frac{1}{2}\ln(Q^+/Q^-)$ is the $\gamma*$ rapidity so that $x_A=e^yQ/\sqrt{s}$
and $x_B=e^{-y}Q/\sqrt{s}$. We have in mind that $Q$ is near $M_Z$ here. In (\ref{css4}), the term involving the $\widetilde{W}_j$
carries the effects from QCD resummation as developed in Refs.~\cite{css}
and the $Y$ term includes those contributions which are 'regular' at $Q_T=p_T \rightarrow 0$ in the sense
of Refs.~\cite{css}, i.e., order by order in perturbation 
theory they are 
derived from the parts of the attendant hard scattering coefficients that are 
less singular than $Q_T^{-2}\times (\text{logs or}\; 1)$ or 
$\delta(\vec{Q_T})$ as $Q_T=p_T \rightarrow 0$. We refer the reader to Refs.~\cite{css} for the remaining notations in (\ref{css4}). 
Our question concerns 
the physical precision of 
the term involving the $\widetilde{W}_j$; for, the $Y$ term is perturbative and can be computed
in principle to the required accuracy by the standard methods. What we note
in the fourth paper in Ref.~\cite{herwiri1} is that the resummed term drops terms ${\cal O}(Q_T/Q)$ in all orders of $\alpha_s$. For example, at $Q_T=5$GeV and $Q=M_Z$ this gives a 5.5\% physical precision error(PPE). We also note~\cite{herwiri1} that the errors on the non-perturbative functions $g_\ell$ yield a $\sim 1.5$\% PPE.
Evidently, this approach to resummation is not precise enough for the 1\% precision tag that we seek with our approach in (\ref{subp15b}); it may be used to give approximations to our new residuals $\tilde{\bar\beta}_{m,n}$ for qualitative studies of consistency, for example. We will address such 
matters elsewhere~\cite{elswh}.
\par
Focussing on the DGLAP-CS theory itself and applying 
the formula in (\ref{subp15b}) to the
calculation of the kernels, $P_{AB}$, we arrive at 
an improved IR limit of these kernels. In this IR-improved DGLAP-CS theory~\cite{irdglap1,irdglap2} large IR effects are resummed for the kernels themselves.
The resulting new resummed kernels, $P^{exp}_{AB}$~\cite{irdglap1,irdglap2}, yield a new resummed scheme for the PDF's and the reduced cross section: 
\begin{equation}
\begin{split}
F_j,\; \hat\sigma &\rightarrow F'_j,\; \hat\sigma'\; \text{for}\nonumber\\
P_{gq}(z)&\rightarrow P^{\text{exp}}_{gq}(z)=C_FF_{YFS}(\gamma_q)e^{\frac{1}{2}\delta_q}\frac{1+(1-z)^2}{z}z^{\gamma_q}, \text{etc.}.
\end{split}
\end{equation}
This new scheme gives the same value for $\sigma$ in (\ref{bscfrla}) with improved MC stability
as discussed in Ref.~\cite{herwiri1}. Here, the YFS~\cite{yfs} infrared factor 
is given by $F_{YFS}(a)=e^{-C_Ea}/\Gamma(1+a)$ where $C_E$ is Euler's constant
and we refer the reader to Ref.~\cite{irdglap1,irdglap2} for the definition of the infrared exponents $\gamma_q,\; \delta_q$ as well as for the complete
set of equations for the new $P^{exp}_{AB}$. $C_F$ is the quadratic Casimir invariant for the quark color representation.\par
The basic physical idea underlying the new kernels,
which was already shown by Bloch and Nordsieck~\cite{bn1}, is illustrated in Fig. 1: 
\begin{figure}[h]
\begin{center}
\epsfig{file=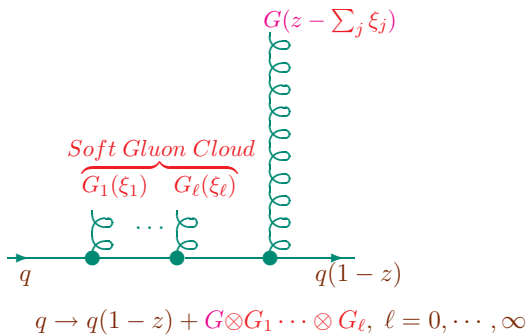,width=70mm}
\end{center}
\label{fig-bn-1}
\caption{Bloch-Nordsieck soft quanta for an accelerated charge.}
\end{figure}
the coherent state of very soft massless quanta of the respective gauge field generated by an accelerated charge makes it impossible to know which of the infinity of possible states
one has made in the splitting process $q(1)\rightarrow q(1-z)+G\otimes G_1\cdots\otimes G_\ell,\; \ell=0,\cdots,\infty$ illustrated in Fig. 1.
The new kernels take this effect into account: they resum terms
${\cal O}((\alpha_s\ln(q^2/\Lambda^2)\ln(1-z))^n)$ for the IR limit $z\rightarrow 1$ to generate the Gribov-Lipatov exponents $\gamma_A$ which therefore start in ${\cal O}(\hbar)$ in the loop expansion. See Refs.~\cite{herwiri1} for
a calculation of the $\gamma_A$.\par
\begin{figure}[h]
\begin{center}
\setlength{\unitlength}{0.1mm}
\begin{picture}(1600, 930)
\put( 370, 770){\makebox(0,0)[cb]{\bf (a)} }
\put(1240, 770){\makebox(0,0)[cb]{\bf (b)} }
\put(   -50, 0){\makebox(0,0)[lb]{\includegraphics[width=80mm]{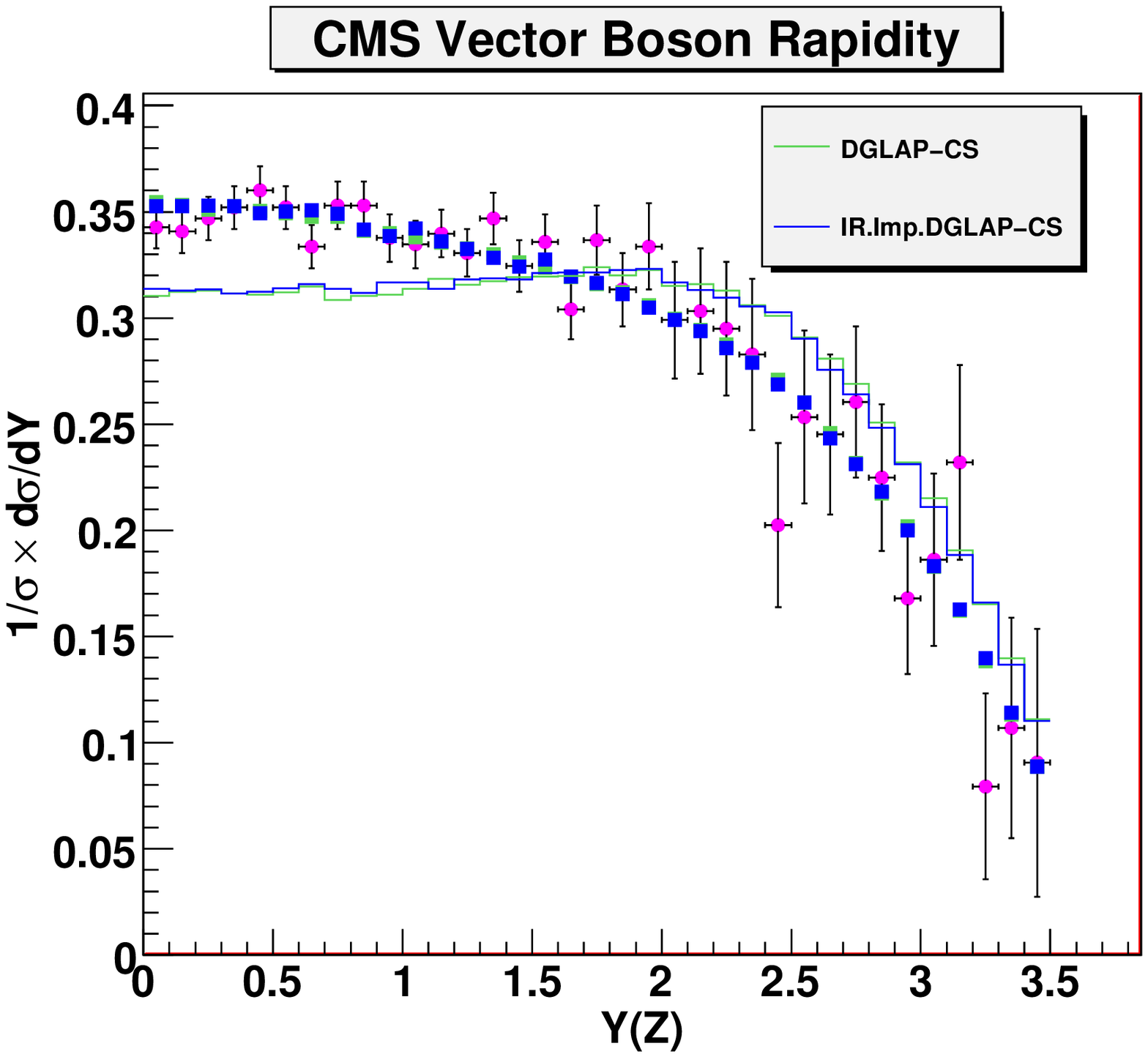}}}
\put( 830, 0){\makebox(0,0)[lb]{\includegraphics[width=80mm]{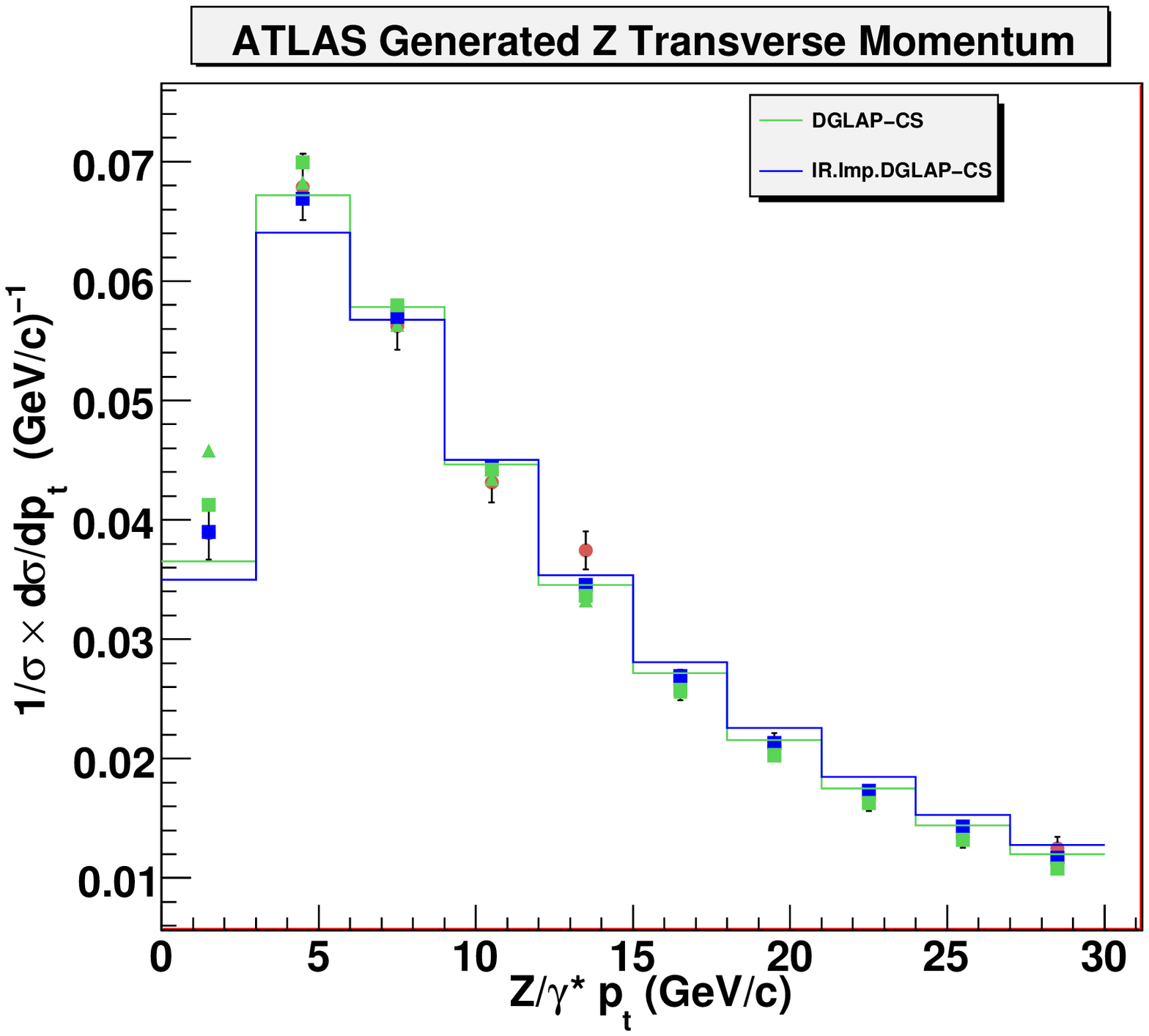}}}
\end{picture}
\end{center}
\caption{\baselineskip=8pt Comparison with LHC data: (a), CMS rapidity data on
($Z/\gamma^*$) production to $e^+e^-,\;\mu^+\mu^-$ pairs, the circular dots are the data, the green(blue) lines are HERWIG6.510(HERWIRI1.031); 
(b), ATLAS $p_T$ spectrum data on ($Z/\gamma^*$) production to (bare) $e^+e^-$ pairs,
the circular dots are the data, the blue(green) lines are HERWIRI1.031(HERWIG6.510). In both (a) and (b) the blue(green) squares are MC@NLO/HERWIRI1.031(HERWIG6.510($\rm{PTRMS}=2.2$GeV)). In (b), the green triangles are MC@NLO/HERWIG6.510($\rm{PTRMS}=$0). These are otherwise untuned theoretical results. 
}
\label{fig2-nlo-iri}
\end{figure}

The new MC Herwiri1.031~\cite{herwiri1} gives the first realization of the new IR-improved kernels in the Herwig6.5~\cite{hrwg} environment. We are in the process of realizing the new kernels in the Herwig++~\cite{hwg++}, Pythia8~\cite{pyth8}, Sherpa~\cite{shrpa} and Powheg~\cite{pwhg} environments as well. Here, we illustrate in Fig.~\ref{fig2-nlo-iri} some of the recent comparisons we have made between Herwiri1.031 and Herwig6.510, 
both with and without
the MC@NLO~\cite{mcnlo} exact ${\cal O}(\alpha_s)$ correction, 
in relation to the LHC data~\cite{cmsrap,atlaspt} on $Z/\gamma*$ production with decay to lepton pairs\footnote{Similar comparisons were made in relation to such data~\cite{d0pt,galea} from FNAL 
in Refs.~\cite{herwiri1} and we comment presently on the connection between
the two sets of comparisons.}.
Just as we found in Refs.~\cite{herwiri1} for the FNAL data on single $Z/\gamma^*$ production, the unimproved MC requires the very hard value of $\text{\rm PTRMS}\cong 2.2$GeV 
to give a good fit to the $p_T$ spectra as well as the rapidity spectra whereas the IR-improved calculation gives very good fits to both of the spectra without the need of such a hard value of {\rm PTRMS}, the rms value for
an intrinsic Gaussian $p_T$ distribution, for the proton wave function: the $\chi^2/d.o.f$ are respectively $(0.72,0.72),\;(1.37,0.70),\;
(2.23,0.70)$ for the $p_T$ and rapidity data for the MC@NLO/HERWIRI1.031,
 MC@NLO/HERWIG6.510($\rm{PTRMS}=2.2$GeV) and MC@NLO/HERWIG6.510($\rm{PTRMS}=$0)
results. Such a hard intrinsic value of {\rm PTRMS} contradicts the results
in Refs.~\cite{rvndl,bj}, as we discuss in Refs.~\cite{herwiri1}.
To illustrate the size of the exact ${\cal O}(\alpha_s)$ correction, we also show the
results for both Herwig6.510(green line) and Herwiri1.031(blue line) without it in the plots in Fig.~\ref{fig2-nlo-iri}. As expected, the exact ${\cal O}(\alpha_s)$ correction is important for both the $p_T$ spectra and the rapidity spectra. The suggested accuracy at the 10\% level shows
the need for the NNLO extension of MC@NLO, in view of our goals
for this process. We also note that, with the 1\% precision goal, one also needs per mille level control of the EW corrections. This issue is addressed in the new version of the ${\cal KK}$ MC~\cite{kkmc422}, version 4.22, which now allows for incoming quark antiquark beams -- see Ref.~\cite{kkmc422} for further discussion of the relevant effects in relation 
to other approaches~\cite{otherEW}.\par 
As we show in Refs.~\cite{herwiri1}, one may use the new precision data at ATLAS
and CMS, where one has now more than $10^7$ $Z/\gamma*$ decays to lepton pairs per experiment, to distinguish between the fundamental description in Herwiri1.031 and the
ad hocly hard intrinsic $p_T$ in Herwig6.5 by comparing the data to the predictions of the detailed line shape and of the more finely binned $p_T$ spectra --
see Figs.~3 and 4 in the last two papers in Refs.~\cite{herwiri1}\footnote{We note that already in Refs.~\cite{pt-comp-LHC} the discriminating power
of $p_T$ spectra in single $Z/\gamma^*$ production at the LHC among theoretical predictions is manifest -- see the last paper in Refs.~\cite{herwiri1} for more discussion on this point.}.
We await the releases of the new precision data accordingly.
\par
\section{Resummed Quantum Gravity: Comparison with Data}
One of us(B.F.L.W.), using his application of exact amplitude-based resummation theory to Feynman's formulation of Einstein's theory, as described in Refs.~\cite{rqg}, has arrived in Ref.~\cite{bw-lambda} at a first principles
prediction of the cosmological constant that is close to the observed value~\cite{cosm1,pdg2008}, $\rho_\Lambda\cong (2.368\times 10^{-3}eV(1\pm 0.023))^4$. 
We now recapitulate this promising result and some of the cross checks
that it has passed.\par
Using the deep UV result
\begin{equation} 
B''_g(k)=\frac{\kappa^2|k^2|}{8\pi^2}\ln\left(\frac{m^2}{m^2+|k^2|}\right),       
\label{yfs1} 
\end{equation}
it is shown in Ref.~\cite{bw-lambda} that the UV limit of Newton's constant, $G_N(k)$, is given by
\begin{equation}
g_*=\lim_{k^2\rightarrow \infty}k^2G_N(k^2)=\frac{360\pi}{c_{2,eff}}\cong 0.0442,
\end{equation} 
where~\cite{rqg,bw-lambda} $c_{2,eff}\cong 2.56\times 10^4$ for the known world.
The same formula (\ref{yfs1}) allows one to show~\cite{bw-lambda} that the contribution of a scalar field to $\Lambda$
is
\begin{equation}
\Lambda_s=-8\pi G_N\frac{\int d^4k}{2(2\pi)^4}\frac{(2k_0^2)e^{-\lambda_c(k^2/(2m^2))\ln(k^2/m^2+1)}}{k^2+m^2}
\cong -8\pi G_N\left[\frac{1}{G_N^{2}64\rho^2}\right],
\label{lambscalar}
\end{equation} 
where $\rho=\ln\frac{2}{\lambda_c}$ for $\lambda_c=\frac{2m^2}{\pi M_{Pl}^2}$ 
and we have used the calculus
of Refs.~\cite{rqg,bw-lambda}. We note that
standard methods~\cite{bw-lambda} then allow one to  
show that a Dirac fermion contributes $-4$ times $\Lambda_s$ to
$\Lambda$, so that 
the deep UV limit of $\Lambda$ becomes
\begin{equation}
\begin{split}
\Lambda(k) &\operatornamewithlimits{\longrightarrow}_{k^2\rightarrow \infty} k^2\lambda_*,\cr
\lambda_*&=-\frac{c_{2,eff}}{2880}\sum_{j}(-1)^{F_j}n_j/\rho_j^2
\cong 0.0817
\end{split}
\end{equation} 
where $F_j$ is the fermion number of $j$ and $\rho_j=\rho(\lambda_c(m_j))$.
Our results for $(g_*,\lambda_*)$ agree qualitatively with those in Refs.~\cite{reutera,reuter1}. Indeed, as we show in Ref.~\cite{bw-lambda}, there is no disagreement in principle between
our gauge invariant, cut-off independent results and the gauge dependent, cut-off dependent results in Refs.~\cite{reutera,reuter1}.
\par
\subsection{A Resummed Quantum Gravity Estimate of $\Lambda$}
Toward obtaining an estimate the value of $\Lambda$ today, we make use of the normal-ordered form of Einstein's equation, 
\begin{equation}
:G_{\mu\nu}:+\Lambda :g_{\mu\nu}:=-8\pi G_N :T_{\mu\nu}:.
\label{eineq2}
\end{equation}
From the coherent state representation of the thermal density matrix one then
arrives at
the Einstein equation in the form of thermally averaged quantities with
$\Lambda$ given by our result above in lowest order. 
Using the result from Refs.~\cite{reuter1} that
the transition time between the Planck regime and the classical Friedmann-Robertson-Walker(FRW) regime is $t_{tr}\sim 25 t_{Pl}$,
we introduce
\begin{equation}
\rho_\Lambda(t_{tr})\equiv\frac{\Lambda(t_{tr})}{8\pi G_N(t_{tr})}
         =\frac{-M_{Pl}^4(k_{tr})}{64}\sum_j\frac{(-1)^Fn_j}{\rho_j^2}
\end{equation}
and follow the arguments in Refs.~\cite{branch-zap} ($t_{eq}$ is the time of matter-radiation equality) to get 
\begin{equation}
\begin{split}
\rho_\Lambda(t_0)&\cong \frac{-M_{Pl}^4(1+c_{2,eff}k_{tr}^2/(360\pi M_{Pl}^2))^2}{64}\sum_j\frac{(-1)^Fn_j}{\rho_j^2}\cr
          &\qquad\quad \times \big[\frac{t_{tr}^2}{t_{eq}^2} \times (\frac{t_{eq}^{2/3}}{t_0^{2/3}})^3\big]\cr
          &\cong \frac{-M_{Pl}^2(1.0362)^2(-9.197\times 10^{-3})}{64}\frac{(25)^2}{t_0^2}\cr
   &\cong (2.400\times 10^{-3}eV)^4.
\end{split}
\label{eqcos1}
\end{equation}
where we take the age of the universe to be $t_0\cong 13.7\times 10^9$ yrs. 
In (\ref{eqcos1}), the first factor in the square bracket comes from the period from
$t_{tr}$ to $t_{eq}$ (radiation dominated) and the second factor
comes from the period from $t_{eq}$ to $t_0$ (matter dominated)
\footnote{The operator field method forces the vacuum energies to follow the same scaling as the non-vacuum excitations.}.
This estimate should be compared with the experimental result~\cite{cosm1,pdg2008}\footnote{See also Ref.~\cite{sola2} for an analysis that suggests a value for $\rho_\Lambda(t_0)$ that is qualitatively similar to this result.}
$\rho_\Lambda(t_0)|_{\text{expt}}\cong (2.368\times 10^{-3}eV(1\pm 0.023))^4$.
\par
In Ref.~\cite{bw-lambda}, it is shown that the result in (\ref{eqcos1}) is robust to the corrections associated with the EW, QCD chiral and GUT suymmetry breaking scales, as these are suppressed by a factor $\sim {\mu_{\text{Breaking}}}^4/(.01 M_{Pl}^4)$
if the respective breaking scale is $\mu_{\text{Breaking}}$. It is also shown in Ref.~\cite{bw-lambda} that continuity of the Hubble parameter across the boundary from the Planck regime to the FRW regime in the model of Ref.~\cite{reuter1}
requires a gauge transformation, which, if taken as a dilatation, shows that 
the result in (\ref{eqcos1}) leads to the value $\Omega_\Lambda(t_{\text{BBN}})\cong 1.31\times 10^{-3}$, so that it does not significantly affect Big Bang Nucleosynthesis(BBN)~\cite{bbn}, where we use standard notation. The presence of possible higher degrees of freedom such as  those in susy GUT models~\cite{ravi-1} is discussed in Ref.~\cite{bw-lambda} with the conclusion that such models are consistent with (\ref{eqcos1}) only if they are modified with new degrees of freedom at scales much higher than the EW scale -- see Ref.~\cite{bw-lambda} for the detailed discussion. Finally, we note~\cite{bw-lambda}, concerning the issue of the covariant conservation of matter in the current universe, that only when $\dot{\Lambda}+8\pi\dot{G}_N=0$ holds is this guaranteed and that violations of such conservation are allowed as long as they are small, as discussed in Refs.~\cite{sola-cons}.\par
In closing, two of us (B.F.L.W., S.A.Y.)
thank Prof. Ignatios Antoniadis for the support and kind 
hospitality of the CERN TH Unit while part of this work was completed.
\par

\end{document}